\documentclass[12pt,a4paper]{article}
\usepackage{graphicx}
\usepackage{color}
\usepackage{bm,enumerate,amsmath,amssymb,amsthm}
\usepackage{epsfig}
\usepackage{cite}
\usepackage{algorithm}
\usepackage{algpseudocode}
\usepackage{txfonts}
\usepackage{subfigmat}
\usepackage{figsize}
\usepackage{here}
\usepackage{listings}
\usepackage{braket}
\usepackage{comment}
\usepackage{lscape}
\usepackage{bigints}

\usepackage{arxiv}
\usepackage[utf8]{inputenc} 

\theoremstyle{definition}

\newtheorem*{theorem*}{Theorem}

\newtheorem*{definition*}{Definition}

\renewcommand{\headeright}{}

\title{Scaling Behavior of Parameterized Quantum Circuits from a Lie-Algebraic Perspective}

\author{
  Hiroshi Ohno\\
  Toyota Central R\&D Labs., Inc.\\
  Aichi, Japan\\
  \texttt{oono-h@mosk.tytlabs.co.jp}\\
}

\date{\empty}

\begin{document}

\maketitle

\begin{abstract}
  Understanding how the performance of parameterized quantum circuits scales with available resources is important for characterizing their trainability and effective model capacity.
  In this study, we numerically investigate data scaling, model scaling, and compute scaling in parameterized quantum circuits and examine Lie-algebraic quantities as alternative measures of model size.
  In addition to the number of circuit parameters, we consider the dimension of the dynamical Lie algebra, the observable-orbit dimension, and a Jacobian effective dimension defined as the rank of the Jacobian of the parameterized observable orbit.
  Using a regression task with randomly generated Pauli-string generators, we observe decreasing loss with increasing training dataset size, parameter size, and number of optimization iterations over the ranges investigated.
  For model scaling, the dynamical Lie algebra and observable orbit dimensions rapidly saturate as the parameter size increases, whereas the Jacobian effective dimension remains strongly correlated with the parameter size and exhibits comparable scaling behavior.
  These results suggest that the Jacobian effective dimension provides a geometry-aware measure of the locally accessible observable degrees of freedom of finite-depth parameterized quantum circuits and may serve as a useful scaling parameter beyond the nominal parameter size.
\end{abstract}

\section{Introduction}
Scaling laws provide empirical relationships between model performance and the resources used for training.
In classical machine learning, performance has been studied as a function of quantities such as training data size, model size, and training compute: $ loss \propto X^{\alpha} $ \cite{kaplan2020}.
Here, $ X $ is called a scaling parameter and $ \alpha $ is called a scaling coefficient in this study.
Such relationships are useful not only for characterizing empirical performance trends but also for understanding how additional resources translate into improved learning performance.

Similar questions arise for parameterized quantum circuits (parameterized unitaries).
The performance of a parameterized unitary may depend on several resource measures, including the amount of training data, the number of trainable parameters (parameter size), circuit depth, and optimization cost.
Recent studies have begun to investigate scaling behavior in quantum and hybrid quantum-classical learning models.
Vyskubov et al. \cite{vyskubov2026} studied the scaling of hybrid quantum neural networks with circuit depth and qubit count, while Haug et al. \cite{haug2021} identified scaling behavior in the expressive capacity of parameterized quantum circuits using an effective quantum dimension.
More recently, scaling-law approaches analogous to those used in classical machine learning have also been applied to neural-network quantum states, where approximation accuracy was found to exhibit power-law behavior with training compute \cite{rende2026}.
However, the number of trainable parameters alone does not necessarily characterize the effective degrees of freedom accessible to a quantum circuit.
Because the generators of a parameterized unitary possess an underlying Lie-algebraic structure, two circuits with similar parameter size may, in principle, explore different subsets of the unitary or observable space.

This observation motivates the use of geometry- and Lie-algebra-based quantities as alternative measures of model size.
In this study, we consider three such quantities: the dimension of the dynamical Lie algebra (DLA), the dimension of the observable orbit (OOD), and the Jacobian effective dimension ($ D_{eff} $).
The DLA characterizes the Lie algebra generated by the generators, while the OOD characterizes the dimension of the observable orbit associated with the generated Lie group.
These quantities describe global properties of the accessible model space.
In contrast, for a finite-depth parameterized ansatz, $ D_{eff} $ measures the number of locally accessible directions of the parameterized observable orbit at a given parameter point.

We numerically investigate three types of scaling: data scaling, model scaling, and compute scaling.
The corresponding conventional scaling parameters are the training sample size $ M $, the parameter size $ P $, and the number of optimization iterations $ T $, respectively.
For model scaling, we additionally examine DLA, OOD, and $ D_{eff} $ as candidate scaling parameters.
Using a regression task and parameterized unitaries generated from randomly sampled Pauli strings, we study how the loss varies with these quantities.

Our numerical results show decreasing loss with increasing $ M $, $ P $, and $ T $ over the parameter ranges considered.
For model scaling, we find that DLA and OOD rapidly approach their maximal values as the number of circuit parameters increases, which limits their dynamic ranges as scaling parameters in the present setting.
By contrast, $ D_{eff} $ remains strongly correlated with $ P $ and yields scaling behavior comparable to that obtained using the parameter size itself.
These observations suggest that $ D_{eff} $ provides an effective, geometry-aware characterization of finite-depth parameterized unitaries by quantifying the observable directions that are locally accessible through parameter variations.

The remainder of this paper is organized as follows.
Section \ref{sec3} introduces the parameterized unitary model and the Lie-algebraic and Jacobian-based quantities considered as scaling parameters.
Section \ref{sec4} describes the numerical experiments and presents the scaling results.
Section \ref{sec5} summarizes our findings and discusses limitations and future directions.


\section{Method}\label{sec3}
We consider the following parameterized unitary (ansatz):
\begin{equation}\label{eq1}
  U(\bm{\theta}) = \prod_{j=1}^{L} e^{-i \theta_{j} G_{j}},
\end{equation}
where $ L $ denotes the circuit depth, $ \bm{\theta} = (\theta_{1}, \ldots, \theta_{L})^{\mathsf{T}} \in \bm{\Theta}_{L} \subset \mathbb{R}^{L} $, and $ G_{j} $ denotes a generator, where $ \mathsf{T} $ denotes transpose.
Here, the parameter size $ P $ is equal to $ L $.

Motivated by the underlying Lie-algebraic structure, we consider the following quantities as candidate model-scaling parameters:
\begin{itemize}
\item Dimension of dynamical Lie algebra (DLA),
\item Dimension of observable orbit (OOD),
\item Jacobian effective dimension ($ D_{eff} $).
\end{itemize}
The ansatz (Eq. \ref{eq1}) generates the DLA as follows:
\begin{equation}\label{eq2}
  \mathfrak{g} = {\rm Lie} (i G_{1}, \ldots, i G_{L}) \subseteq \mathfrak{u}(N),
\end{equation}
where $ N = 2^{n} $ and $ n $ is the number of qubits.
The corresponding Lie group $ G $ is $ \exp(\mathfrak{g}) \subseteq {\rm U}(N)$.

OOD is defined as $ {\rm dim}(\{[X, \mathcal{O}] \mid X \in \mathfrak{g} \}) $, which corresponds to the orbit dimension of the class of all Lie groups $ G $, where $ \mathcal{O} $ denotes an observable.
OOD can be regarded as the maximum capacity of the model class.
When $ \mathfrak{g} $ and $ \mathcal{O} $ are fixed, OOD becomes a constant, thus cannot be exploited as a scaling parameter.
Since $ L $ is finite, the ansatz does not correspond to the class of all Lie groups $ G $.
Thus, we define the following set:
\begin{equation}\label{eq3}
  O_{\mathcal{O}, L} \coloneqq \{ U(\bm{\theta})^{\dagger} \mathcal{O} U(\bm{\theta}) \}.
\end{equation}
This set represents the portion of the observable orbit accessible to the finite-depth ansatz.
Then, we define $ D_{eff} $ as follows:
\begin{equation}\label{eq3}
  D_{eff} = {\rm rank} \left[\frac{\partial {\rm vec}(U^{\dagger} \mathcal{O} U )}{\partial \theta_{1}}, \ldots, \frac{\partial {\rm vec}(U^{\dagger} \mathcal{O} U )}{\partial \theta_{P}} \right].
\end{equation}
During training, $ D_{eff} $ represents the number of locally accessible directions in the observable orbit.
In this study, $ \bm{\theta} $ is used after training.
For $ D_{eff} $, $ P $, and OOD, $ D_{eff} \leq \min(P, {\rm OOD}) $.
As $ P $ increases, $ D_{eff} $ may approach OOD when the parameterization becomes locally expressive enough to span the observable orbit.

The scaling parameters in this study are summarized in Table \ref{tab3-1}.
\begin{table}[htb]
  \centering
  \caption{Scaling parameters}\label{tab3-1}
  \vspace{5pt}
  \begin{tabular}{cc}\hline
    Data scaling & $ M $ (training sample size)\\
    Model scaling & $ P $ (parameter size)\\
    & DLA\\
    & OOD\\
    & $ D_{eff} $\\
    Compute scaling & $ T $ (number of SPSA iterations)\\ \hline
  \end{tabular}
\end{table}
Here, we use the number of SPSA iterations as a proxy for training compute.

\section{Numerical experiments and results}\label{sec4}
\noindent
{\bf Training data preparation:} We addressed a linear regression task and generated training data samples as follows.
The target model was defined as $ y = \beta^{\mathsf{T}} \bm{x} + \epsilon $.
The dimension of $ \bm{x} $ corresponds to the number of qubits $ n $, in this study, $ n = 4 $ and $ n = 8 $ were used.
The coefficient vector $ \beta $ was sampled uniformly from $ [-1, 1] $.
The noise $ \epsilon $ was sampled from the normal distribution with mean zero and standard deviation one.
The input data $ \bm{x} $ was sampled uniformly from $ [0, 2 \pi] $.
The target values $ y $ were linearly normalized to the range $ [-1, 1] $ so that they were compatible with the output range of the quantum model.
To obtain $ M $ samples, the input data and the noise were sampled $ M $ times.\\

\noindent
{\bf Training model:} The output state of a quantum circuit was obtained as $ \ket{\phi} = \left( \prod_{j=1}^{L} \exp(-i \theta_{j} G_{j}) \right) \, RY(x)^{\otimes n} \, \ket{0}^{\otimes n} $.
The corresponding predicted value was obtained as $ \hat{y} = \braket{\phi | \, Z(0) \, | \phi} $.
Each generator $ G_{j} $ was sampled from the set of $ n $-qubit Pauli strings, excluding the all-identity string.

The training loss was measured using the root mean square error (RMSE).
The optimizer was the simultaneous perturbation stochastic approximation (SPSA) \cite{gacon2021}.
The learning rate was 0.001, and the momentum term was 0.5.
The training was performed 10 times with different random seeds.\\

\begin{figure}[htbp]
  \centering
  \begin{minipage}{13.5cm}
    \SetFigLayout{1}{2}
    \subfigure[$ n = 4 $]{\includegraphics[width=6cm]{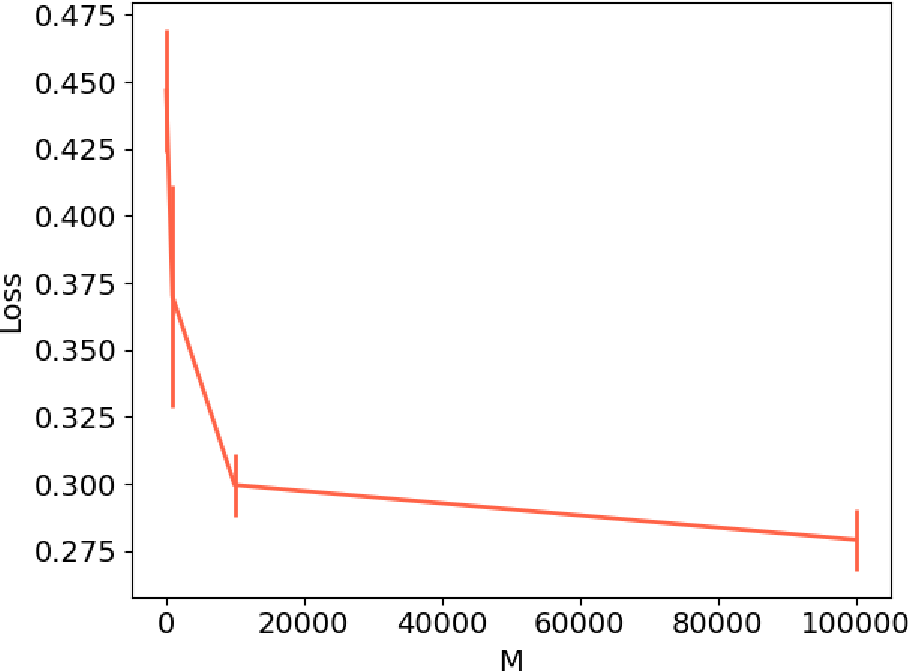}}
    \hfill
    \subfigure[$ n = 8 $]{\includegraphics[width=6cm]{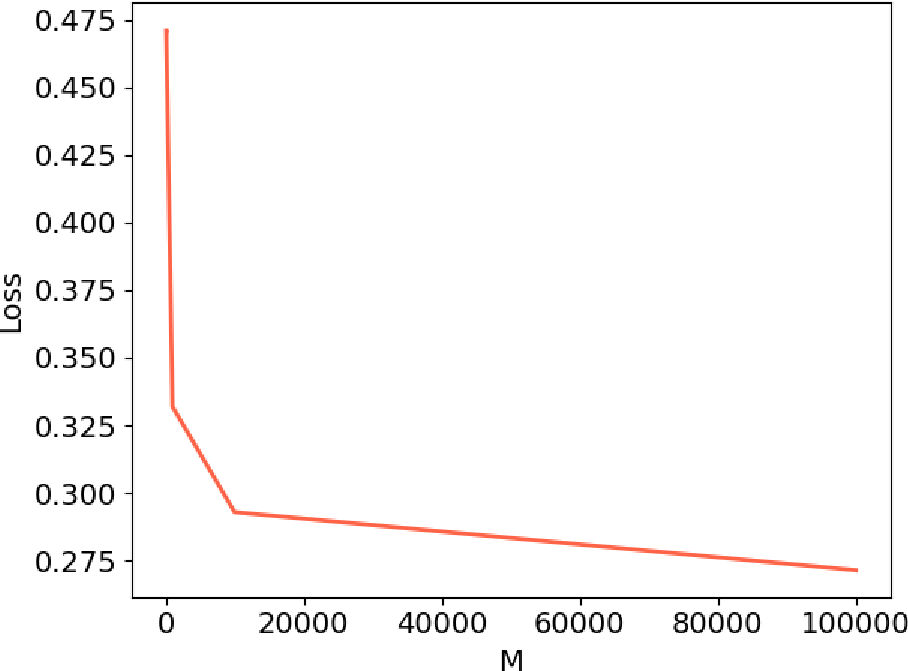}}
  \end{minipage}
  \caption{Loss curves of sample size $ M $. Mean and standard deviation (error bars) of loss are shown.}\label{fig4-1}
\end{figure}

\noindent
{\bf Results of data scaling:} Figure~\ref{fig4-1} shows the average loss results of sample size $ M \in \{ 100, 1000, 10000, 100000 \} $.
The error bars indicate the standard deviation.
Here, $ L = 5 $ ($ P = 5 $), the batch size ($ B $) was 1000, and $ T = 5000 $.
From the log-log fitting, we obtained a scaling coefficient (slope) $ \alpha = -0.07018 $ (0.00446) (the p-value $ < 0.01 $) for $ n = 4 $ and $ \alpha = -0.07732 $ (0.00016) (the p-value $ < 0.01 $) for $ n = 8 $.
Values in parentheses denote standard deviations.
Here, we tested whether the mean slope across ten independent runs was smaller than zero using a one-tailed one-sample t-test.
$ \alpha $ of $ M $ was significantly negative across independent runs.
Therefore, these results support $ M $ as a candidate scaling parameter.
Additionally, DLA was 15 (3.68782), OOD was 7.2 (1.88680), and $ D_{eff} $ was 3.5 (0.80623).\\

\begin{figure}[htbp]
  \centering
  \begin{tabular}{c}
    \includegraphics[width=10cm]{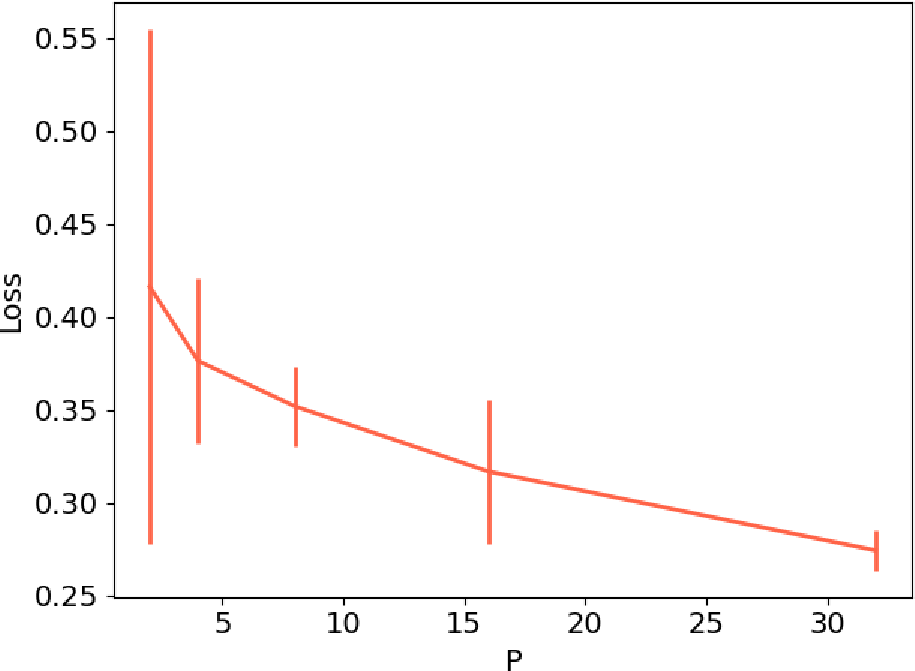}
  \end{tabular}
  \caption{Loss curves of parameter size $ P $. Mean and standard deviation (error bars) of loss are shown.}\label{fig4-2}
\end{figure}

\noindent
{\bf Results of model scaling:} Figure~\ref{fig4-2} shows the average loss results of parameter size $ P \in \{ 2, 4, 8, 16, 32 \} $.
Here, $ n = 4 $, $ M = 1000 $, $ B = 100 $, and $ T = 5000 $.
The scaling coefficient $ \alpha $ was -0.13463 (0.08411) (the p-value $ < 0.01 $).
Therefore, these results support $ P $ as a candidate scaling parameter.

\begin{figure}[htbp]
  \centering
  \begin{minipage}{13.5cm}
    \SetFigLayout{2}{2}
    \subfigure[DLA]{\includegraphics[width=6cm]{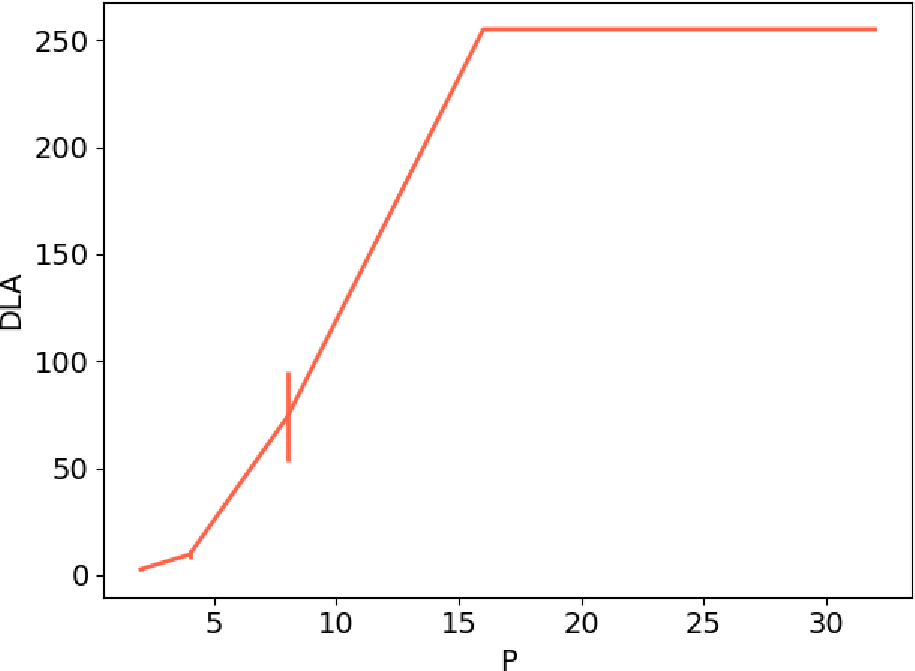}}
    \hfill
    \subfigure[OOD]{\includegraphics[width=6cm]{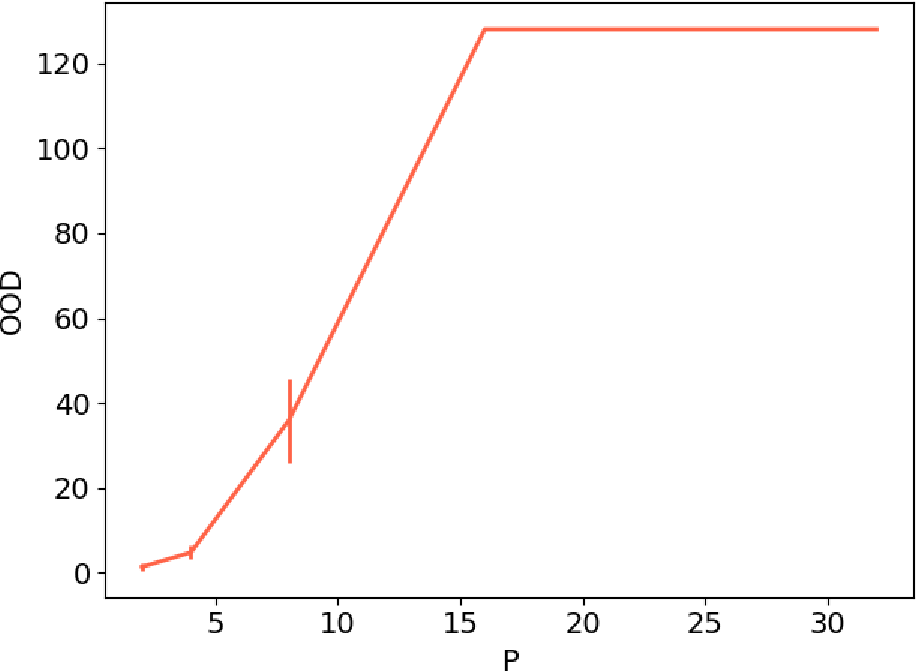}}\\
    \centering
    \subfigure[$D_{eff}$]{\includegraphics[width=6cm]{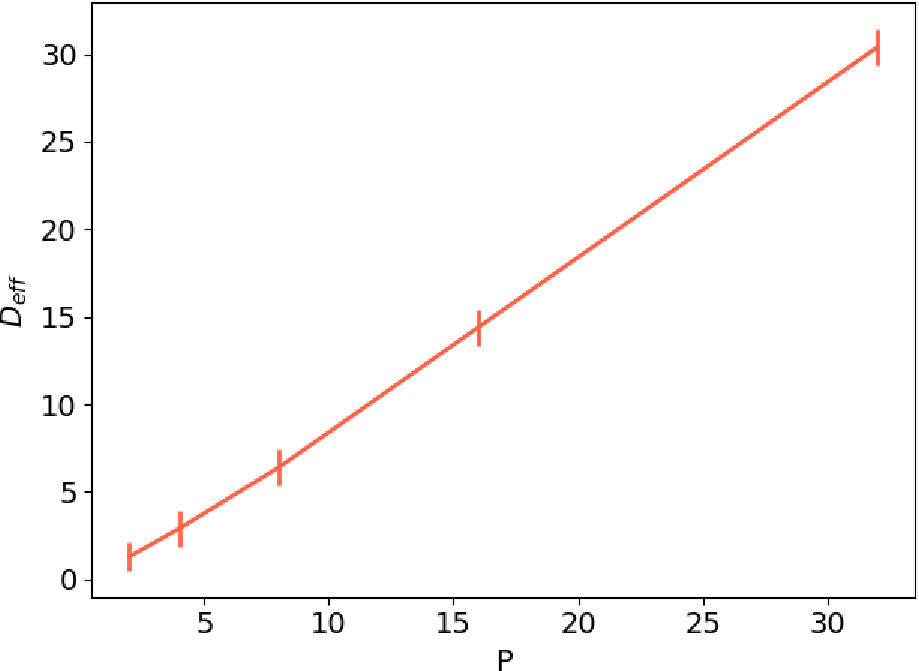}}\\
  \end{minipage}
  \caption{Results of dimension of DLA, OOD, and $ D_{eff} $ with respect to parameter size $ P $. Mean and standard deviation (error bars) are shown.}\label{fig4-3}
\end{figure}

\begin{figure}[htbp]
  \centering
  \begin{minipage}{13.5cm}
    \SetFigLayout{2}{2}
    \subfigure[DLA]{\includegraphics[width=6cm]{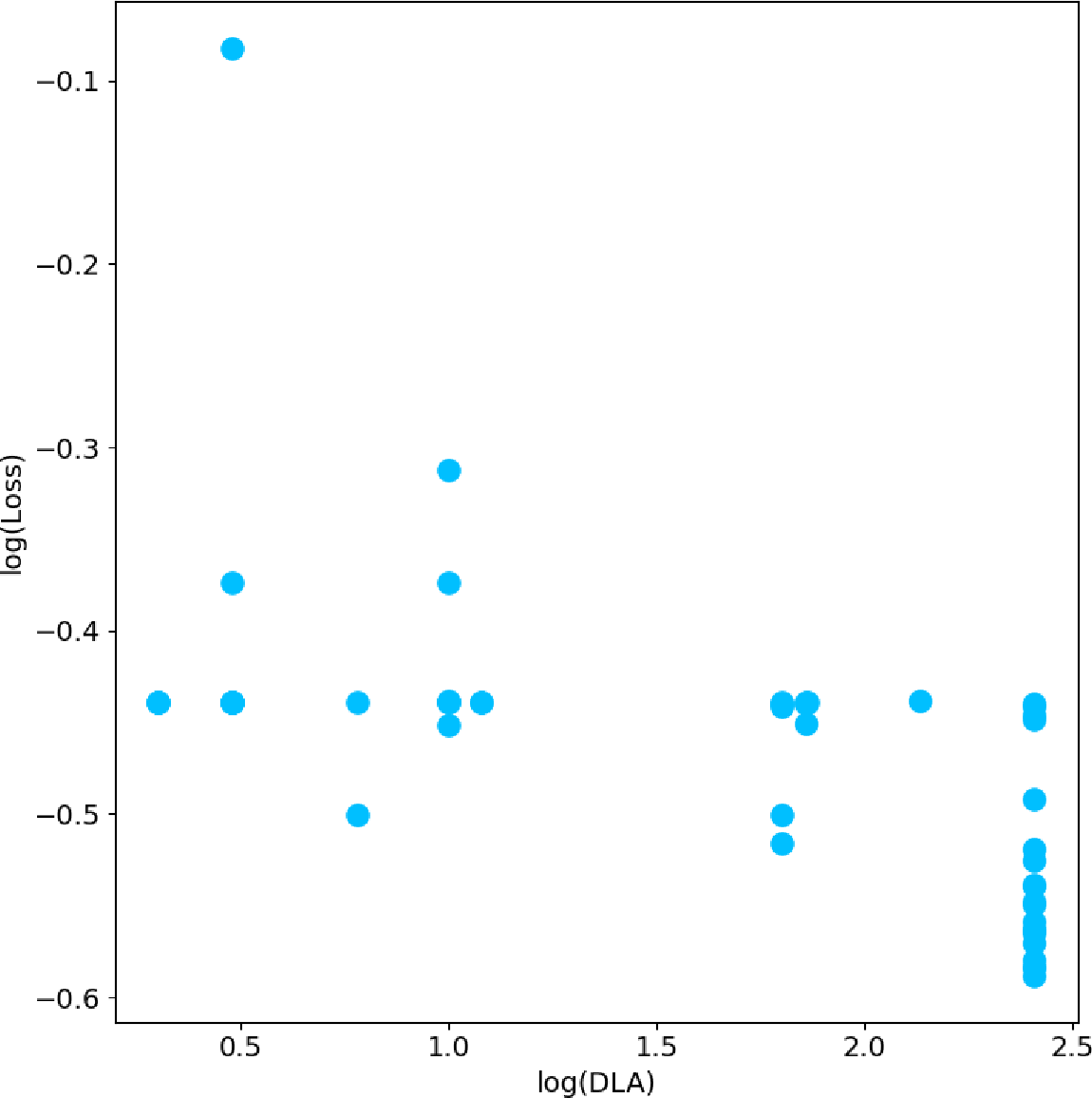}}
    \hfill
    \subfigure[OOD]{\includegraphics[width=6cm]{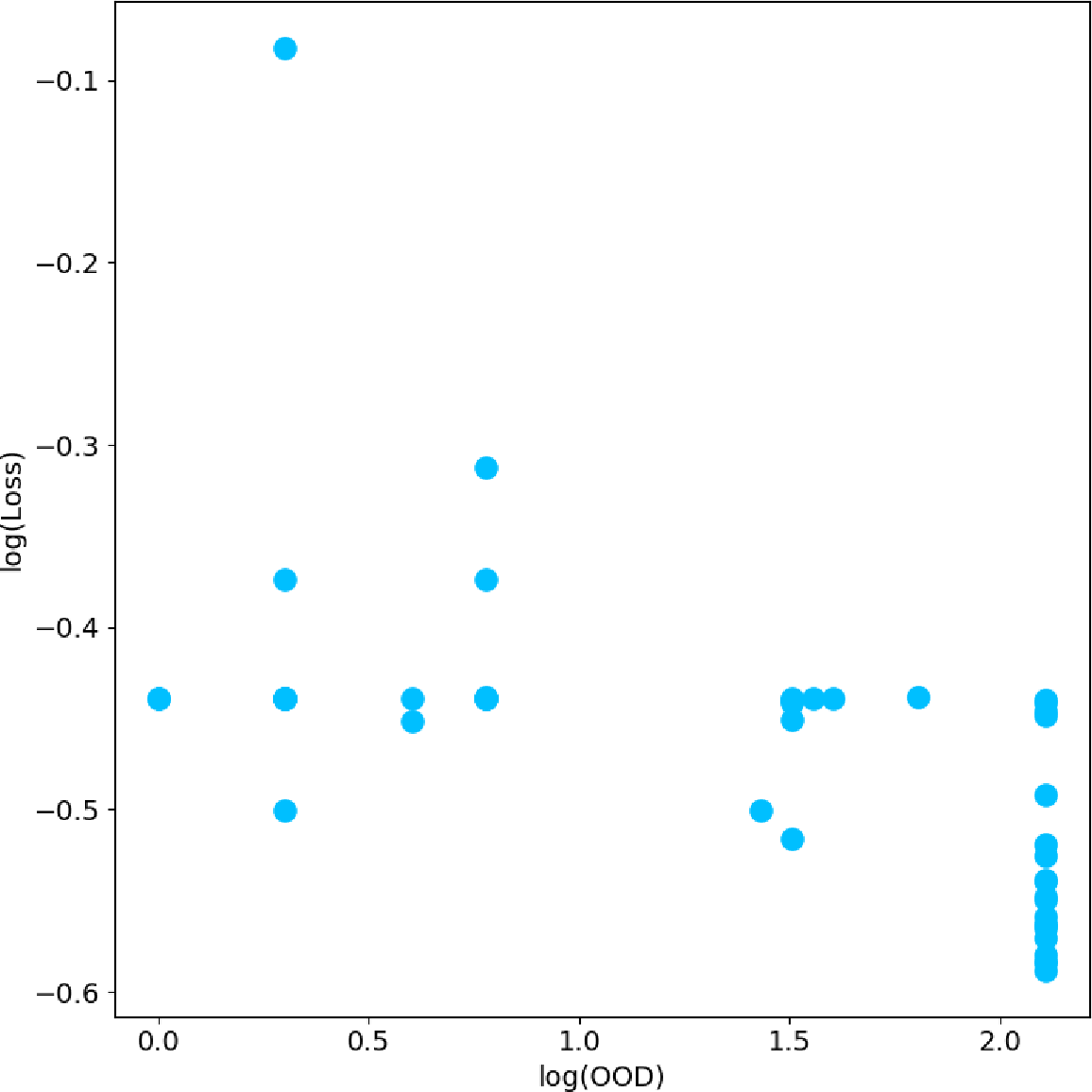}}\\
    \centering
    \subfigure[$D_{eff}$]{\includegraphics[width=6cm]{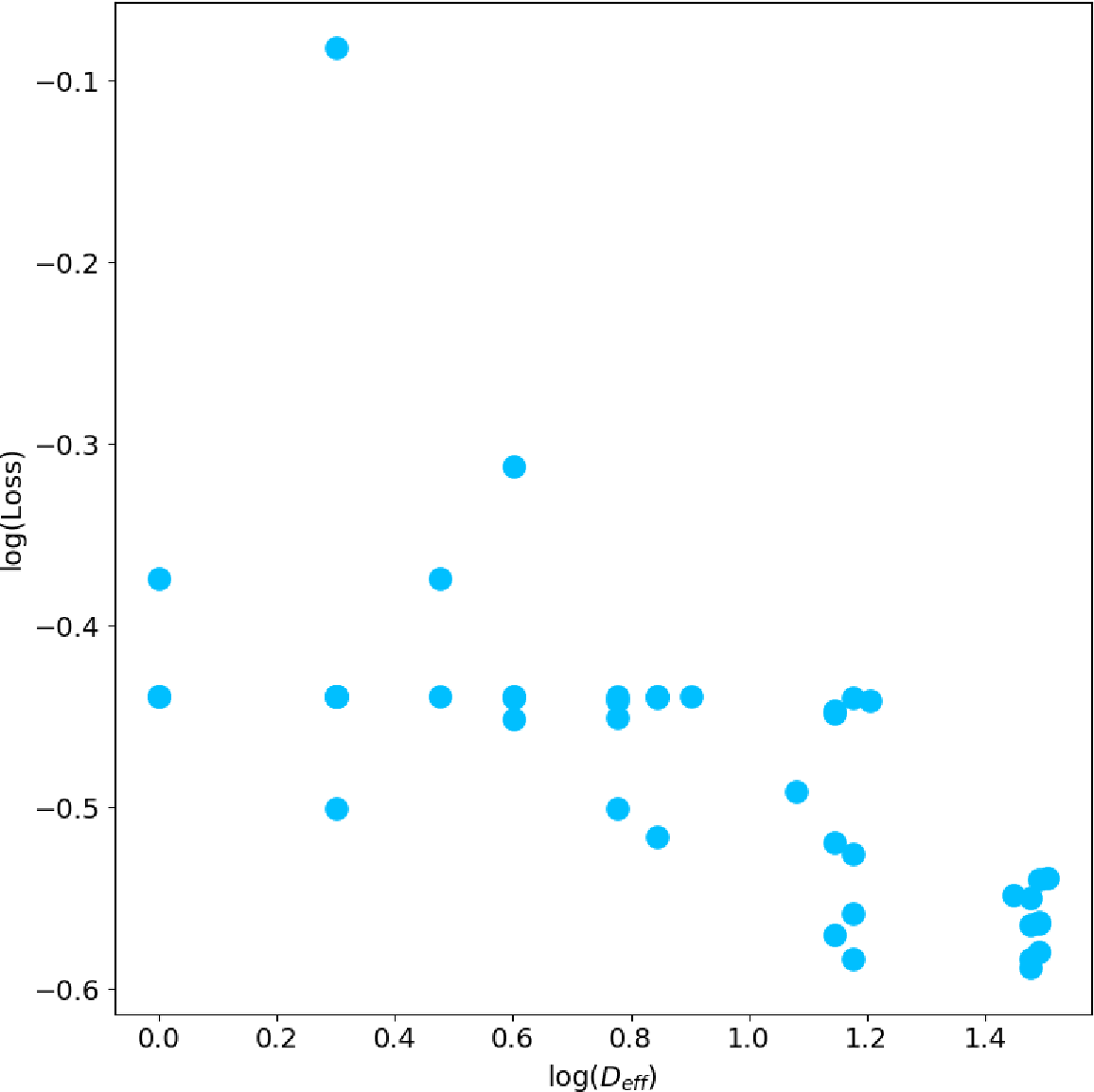}}\\
  \end{minipage}
  \caption{Log-log plots between loss and DLA, OOD, and $ D_{eff} $}\label{fig4-4}
\end{figure}

Figure~\ref{fig4-3} shows the average values of DLA, OOD, and $ D_{eff} $ with respect to the parameter size $ P $.
Figure~\ref{fig4-4} shows the log-log plots between loss and DLA, OOD, and $ D_{eff} $.
DLA and OOD saturated at 255 and 128 for $ P = 16, 32 $, respectively.
The corresponding saturation is also visible in the DLA- and OOD-based plots in Figure \ref{fig4-4}.
This suggests that the system achieves full controllability.
The saturated value $ 255 = 4^{4} - 1 = {\rm dim} \, \mathfrak{su}(16) $ indicates controllability up to a global phase.
$ D_{eff} $ was strongly correlated with $ P $ (Pearson product-moment correlation coefficient: 0.9997 (the p-value $ < 0.01 $)).
Additionally, the $ \alpha $ values were -0.0669 (0.04879) (the p-value $ < 0.01 $) for DLA, -0.07404 (0.05188) (the p-value $ < 0.01 $) for OOD, and -0.12573 (0.08835) (the p-value $ < 0.01 $) for $ D_{eff} $, respectively.
These results indicate that the scaling behavior obtained using $ P $ was nearly identical to that obtained using $ D_{eff} $.
DLA and OOD rapidly saturated for the ansatz considered here, limiting their usefulness as scaling variables in this regime.
Because multiple values of $ P $ map to the same saturated DLA or OOD, the fitted slopes in these coordinates are difficult to interpret as genuine scaling exponents.\\

\begin{figure}[htbp]
  \centering
  \begin{tabular}{c}
    \includegraphics[width=10cm]{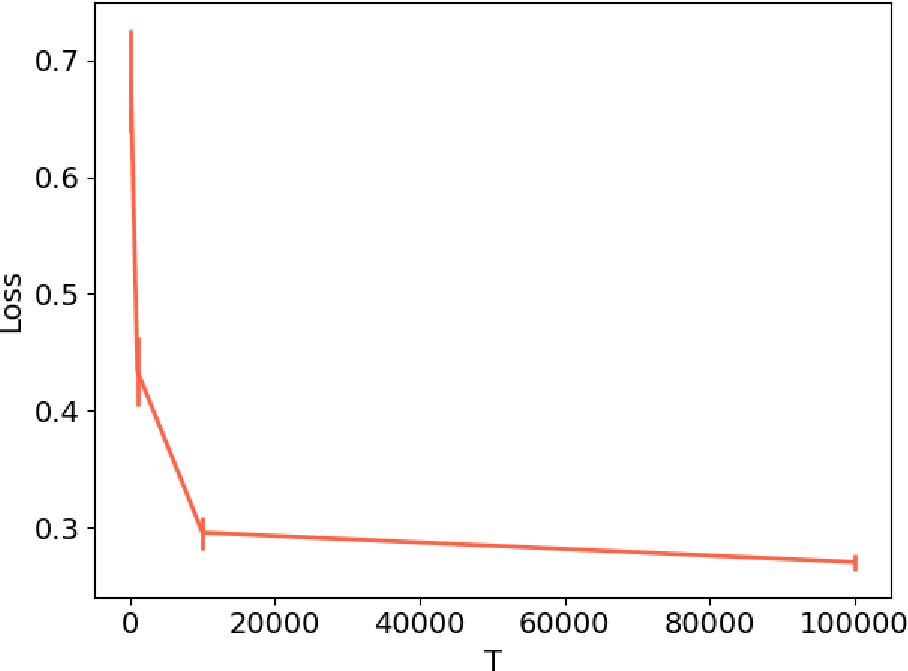}
  \end{tabular}
  \caption{Loss curves of the number of SPSA iterations $ T $. Mean and standard deviation (error bars) of loss are shown.}\label{fig4-5}
\end{figure}

\noindent
{\bf Results of compute scaling:} Figure~\ref{fig4-5} shows the average loss results of the number of SPSA iterations $ T \in \{ 100, 1000, 10000, 100000 \} $.
Here, $ n = 4 $, $ L = 32 $, $ M = 100 $, and $ B = 100 $.
The scaling coefficient $ \alpha $ was -0.13702 (0.01273) (the p-value $ < 0.01 $).
Therefore, these results support $ T $ as a candidate scaling parameter.

\section{Conclusion}\label{sec5}
In this study, we numerically investigated data, model, and compute scaling in parameterized unitaries.
Using a linear regression task, we examined the dependence of the loss on the training sample size $ M $, the parameter size $ P $, and the number of SPSA iterations $ T $.
Over the ranges investigated, the loss decreased as each of these resource parameters increased, and the fitted scaling coefficients ($ \alpha $ values) were consistently negative across independent runs.

For model scaling, we further investigated three quantities motivated by the Lie-algebraic and geometric structure of the parameterized unitary: the dimension of the dynamical Lie algebra (DLA), the observable orbit dimension (OOD), and the Jacobian effective dimension ($ D_{eff} $).
In the random-Pauli ansatz considered in this study, the DLA and OOD rapidly saturated as the parameter size increased.
Consequently, their usefulness as scaling parameters was limited in the saturated regime.
This does not exclude the possibility that these quantities may provide useful scaling parameters in architectures or parameter regimes in which they remain below their maximal values.

In contrast, $ D_{eff} $ remained strongly correlated with $ P $ and produced loss-scaling behavior comparable to that obtained using $ P $.
Unlike the nominal parameter size, $ D_{eff} $ characterizes the number of locally accessible directions of the observable orbit induced by variations of the circuit parameters.
Therefore, it provides a geometry-aware measure of the effective model size of a finite-depth parameterized unitary.

The present study has several limitations.
First, the numerical experiments were restricted to a regression task and relatively small numbers of qubits (four and eight).
Second, the observed scaling behavior was investigated over finite ranges of data size, model size, and optimization iterations; establishing a universal asymptotic power law would require substantially broader numerical studies and comparisons with alternative functional forms.
Third, computing $ D_{eff} $ requires evaluating and rank-testing a Jacobian whose size increases with both the Hilbert-space dimension and the parameter size, which may become computationally expensive for large systems.

An important direction for future work is therefore to investigate whether the relationship between loss and $ D_{eff} $ persists across different learning tasks, circuit architectures, observables, and larger system sizes.
Although speculative, toward realizing large-scale systems, including prospective quantum large language models, quantum circuits have to be large in terms of both the number of qubits and circuit depth.
To mitigate barren plateaus in large circuits, we believe that repetitive architectures consisting of alternating parameterized-unitary layers and measurement operations, whether on all qubits or subsets of qubits, constitute a promising framework with rich expressibility, strong nonlinearity, and mixture-of-experts-like behavior.
As pointed out in \cite{deshpande2025}, due to measurements inserted into part of the circuit, barren plateaus may be avoided.
It will be necessary to extend $ D_{eff} $ to repetitive-architecture models.
Additionally, it will be useful to study the evolution of $ D_{eff} $ during training and to compare its value at initialization with its value after training.
More generally, combining global Lie-algebraic quantities with local Jacobian-based quantities may provide a systematic framework for characterizing the effective capacity and scaling behavior of parameterized unitaries.

\bibliographystyle{plain}
\bibliography{references}

\begin{thebibliography}{1}

\bibitem{deshpande2025}
Abhinav Deshpande, Marcel Hinsche, Khadijeh Najafi, Kunal Sharma, Ryan Sweke,
  and Christa Zoufal.
\newblock Dynamic parameterized quantum circuits: expressive and barren-plateau
  free.
\newblock {\em arXiv preprint arXiv:2411.05760}, 2025.

\bibitem{gacon2021}
Julien Gacon, Christa Zoufal, Giuseppe Carleo, and Stefan Woerner.
\newblock Simultaneous {P}erturbation {S}tochastic {A}pproximation of the
  {Q}uantum {F}isher {I}nformation.
\newblock {\em {Quantum}}, 5:567, Oct 2021.

\bibitem{haug2021}
Tobias Haug, Kishor Bharti, and M.S. Kim.
\newblock Capacity and quantum geometry of parametrized quantum circuits.
\newblock {\em PRX Quantum}, 2:040309, Oct 2021.

\bibitem{kaplan2020}
Jared Kaplan, Sam McCandlish, Tom Henighan, Tom~B. Brown, Benjamin Chess, Rewon
  Child, Scott Gray, Alec Radford, Jeffrey Wu, and Dario Amodei.
\newblock Scaling laws for neural language models.
\newblock {\em arXiv preprint arXiv:2001.08361}, 2020.

\bibitem{rende2026}
Riccardo Rende, Alessandro Sinibaldi, Luciano~Loris Viteritti, Roeland
  Wiersema, Antoine Georges, and Giuseppe Carleo.
\newblock Scaling laws for neural-network quantum states.
\newblock {\em arXiv preprint arXiv:2606.02794}, 2026.

\bibitem{vyskubov2026}
Danil Vyskubov, Kirill Vyskubov, Nouhaila Innan, and Muhammad Shafique.
\newblock Scaling laws for hybrid quantum neural networks: Depth, width, and
  quantum-centric diagnostics.
\newblock {\em arXiv preprint arXiv:2604.06007}, 2026.

\end{thebibliography}

\end{document}